\documentclass[letterpaper]{article} % DO NOT CHANGE THIS
\usepackage[]{aaai2026}  % DO NOT CHANGE THIS
\usepackage{times}  % DO NOT CHANGE THIS
\usepackage{helvet}  % DO NOT CHANGE THIS
\usepackage{courier}  % DO NOT CHANGE THIS
\usepackage[hyphens]{url}  % DO NOT CHANGE THIS
\usepackage{graphicx} % DO NOT CHANGE THIS
\usepackage{natbib}  % DO NOT CHANGE THIS AND DO NOT ADD ANY OPTIONS TO IT
\usepackage{xurl}
\usepackage{caption} % DO NOT CHANGE THIS AND DO NOT ADD ANY OPTIONS TO IT
\usepackage{booktabs}
\usepackage{fancyvrb}
\usepackage{listings}
\usepackage{algorithm}
\usepackage{algorithmic}
\usepackage{xcolor}

\usepackage{newfloat}
\usepackage{listings}
\DeclareCaptionStyle{ruled}{labelfont=normalfont,labelsep=colon,strut=off} % DO NOT CHANGE THIS
\floatstyle{ruled}
\newfloat{listing}{tb}{lst}{}
\floatname{listing}{Listing}
\title{The First Mass Protest on Threads: Multimodal Mobilization and \\ AI-Generated Visuals in Taiwan’s Bluebird Movement}
\author {
    Tracy Weener\textsuperscript{\rm 1},
    Ho-Chun Herbert Chang\textsuperscript{\rm 1}
}
\affiliations {
    \textsuperscript{\rm 1}Program in Quantitative Social Science, Dartmouth College, Hanover, NH 03755 USA
}
\usepackage{bibentry}
\begin{document}

\maketitle

\begin{abstract}
The 2024 Bluebird Movement in Taiwan marked one of the largest youth-led protests in the country’s democratic history, mobilizing over 100,000 demonstrators in response to parliamentary reforms. Unlike the 2014 Sunflower Movement, Bluebird unfolded within a transformed digital environment dominated by Threads, Meta’s new microblogging platform that—uniquely—draws 24\% of its global traffic from Taiwan. Leveraging a dataset of 62,321 posts and 21,572 images, this study analyzes how protest communication developed across textual and visual modalities. We combine LLM zero-shot annotation, gradient-boosting trees, and SHAP explainers, to disambiguate the supply and demand of attention.
Results reveal three dynamics: (1) partisan asymmetries between algorithmic exposure and user endorsement, with anti-DPP content surfaced more widely but anti-KMT and pro-DPP content more actively recirculated; (2) textual repertoires centered on commemorations, personal testimonies, and calls to action as key drivers of virality; and (3) a bifurcation in visual strategies, where human photographs concentrated exposure and discussion, while AI-generated animal and plant symbols circulated as mobilization tools and partisan attacks. These findings demonstrate how Threads functioned as both an amplifier and filter of democratic contention, extending theories of emotional and visual contagion by showing how generative AI reshapes symbolic repertoires in contemporary protest through what we term kawaii toxicity—political attacks cloaked in aesthetics of cuteness.
\end{abstract}


\section{Introduction}

In May 2024, more than 100,000 demonstrators took to the streets of Taipei to protest parliamentary reforms that opponents argued undermined checks and balances in Taiwan's Legislative Yuan~\cite{wang2024protesters}. Triggered by legislative maneuvers expanding investigatory powers in ways that would concentrate authority in the Kuomintang (KMT) and Taiwan People's Party (TPP) coalition, the protests evolved into the Qing Niao (Blue Bird) Movement, one of the largest recall campaigns in modern democracies, with more than 30 legislators facing potential removal from office. As the only Mandarin-speaking democracy, Taiwan has witnessed significant protests before---notably the 2014 Sunflower Movement---but Bluebird exemplifies critical shifts in how democratic contention unfolds: mass migration to a new public platform, deployment of AI-generated visuals at scale, and new paradigms in platform-driven censorship.

Shortly before the movement emerged, Meta launched Threads in July 2023, positioning it globally as a competitor to Twitter (now X). While Threads experienced modest adoption in most markets, it saw extraordinary growth in Taiwan. By April 2025, Taiwan accounted for nearly a quarter of Threads' global web traffic, making it a uniquely important site for public discourse and civic organization~\cite{meta2023threads}. Threads filled a vacuum that other popular platforms like LINE (closed messaging) and Instagram (network-structure) could not serve. It became the "Twitter Taiwan never had," distributing attention across multiple demographics and enabling the kind of real-time public discourse that had been absent from Taiwan's digital ecosystem.

During the Bluebird protests, Threads became the primary space for coordination and debate. But a fundamental question emerges: what happens if a platform’s algorithm privileges content that is different than what protestors want themselves? As early as February 2024, Meta platforms announced it would taper back political content surfaced to its users~\cite{instagram2024politicalcontent}. Knowing this, activists exploited Mandarin's morpho-syllabic properties, where the name "Qing Niao" bore both visual and auditory similarity to "Qing Dao," the protest location. This linguistic adaptation circumvented not state censorship but platform moderation. Fundamentally, activists confronted a divergence between what the platform's algorithm surfaced and what users chose to amplify---what we characterize as an algorithmic exposure and social engagement gap. 

Leveraging 62,321 posts and 21,572 images from Threads between May 2024 and June 2025, we analyze how this first massive protest unfolded on Threads. We focus on three interrelated dynamics: the relationship between algorithmic visibility and user-driven circulation, the textual and visual strategies that secure attention, and the role of AI-generated imagery in mobilization and partisan attack. The proliferation of AI-generated animal symbols—blue birds, winter deer, toads, goblins---introduces what we term \textbf{cute toxicity} (or kawaii toxicity): political attacks cloaked in aesthetics of cuteness. Building on scholarship examining emotional and visual contagion in social movements~\cite{gerbaudo2012tweets,mcadam2001dynamics}, we show how generative AI reshapes symbolic repertoires by packaging partisan animosity in adorable aesthetics~\cite{allison2006millennial,ngai2012our}, with implications extending well beyond Taiwan.

\section{Literature Review}

\subsection{Platform-Mediated Protest}

Social movement scholarship has long emphasized the role of visibility in mobilization. Classical theories describe cycles of contention—waves of protest that rise, diffuse, and decline as challengers and authorities interact~\cite{mcadam2001dynamics}. Within these cycles, focusing events trigger moral shock, converting private grievances into public action~\cite{jasper1998emotions}. The underlying assumption is straightforward: if movements can broadcast these events and messages widely enough, they can mobilize support. Visibility meant power.

The advent of social media appeared to amplify this visibility-power relationship. Platforms like Twitter and Facebook lowered broadcasting costs, especially the rapid dissemination of moral-emotional content~\cite{brady2017emotion,tufekci2017twitter}. However, platforms are not neutral amplifiers but active curators~\cite{gillespie2018custodians}. Contemporary platforms employ algorithmic ranking systems that selectively surface content based on predicted engagement, user behavior, and platform policy objectives~\cite{bozdag2013bias}. These systems intervene between what is posted and what is seen, creating a "marketplace of attention" where algorithms work with networks to shape exposure~\cite{webster2014marketplace}.

This curation creates a fundamental gap between what is posted (supply), what algorithms show (exposure), and what users circulate (engagement)~\cite{bandy2021more}. Rather than uniformly amplifying movements, platforms fragment them into overlapping information environments. This divergence is not a new problem. Researchers have distinguished between what elites supply and what publics demand.~\citet{munger2022rightwing} showed that YouTube's recommendation algorithm shapes which political content gains exposure (supply-side) versus which content audiences actively seek and engage with (demand-side). Their analysis reveals that supply and demand operate according to different rationales: algorithms optimize for watch time and engagement, while users optimize for identity reinforcement and information seeking. 

Importantly, exposure doesn't simply reflect aggregated user preferences.~\citet{gillespie2018custodians} argues platforms make editorial decisions—what to surface, what to suppress, what to monetize—that shape public discourse. These decisions create an "algorithmic public" that may diverge from the "social public," or the conversation users would create through their own circulation choices.
Asymmetric use also exists in user bases.~\citet{freelon2020false} find systematic differences in how political groups engage with platforms: right-leaning activists favor high-volume posting and strategic coordination, while left-leaning activists favor spontaneous individual expression. These asymmetries mean engagement patterns reflect not just message resonance but also group-level strategic differences and platform-specific affordances. 

\subsection{Multimodal Communication and AI-Generated Content}
While textual communication has received substantial scholarly attention, recently, scholars have documented the centrality of visual communication in contemporary movements. The 2020 George Floyd protests demonstrated how a single video could trigger global mobilization, functioning as visceral moral shock that text alone could not achieve~\cite{chang2022justiceforgeorgefloyd}. This visual turn reflects platforms' increasing emphasis on image and video content over text, as well as publics' preferences for emotionally immediate content~\cite{bonilla2015ferguson}. Memes, in particular, have emerged as critical protest infrastructure, which are defined as units of cultural information that spread through imitation and variation~\cite{chang2026generative}.

Generative AI represents a phase change in symbolic production costs. Text-to-image models enable anyone with a smartphone to produce high-quality images in seconds at close-to zero marginal cost. This democratization produces broader participation in aesthetic production, rapid iteration, and proliferation of symbols that serve coordinating functions.

A growing body of work examine AI-generated images on dimensions such as hyperbole, realism, and absurdity. A critical aesthetic dimension is kawaii—Japanese for "cute" theorized as politically significant. Cute aesthetics in Japanese popular culture enable soft power projection and identity formation. Crucially, cuteness creates complex affective responses—simultaneously evoking vulnerability and manipulation~\cite{ngai2012our} by making exploitation appear benign~\cite{miller2017cute}. Therefore, partisan animosity and toxicity is an underexplored area in political campaigning and social movements. 

\subsection{Protests in Chinese: Censorship Evasion and Linguistic Innovation}

Research on the evasion of censorship often occurs under authoritarian contexts.~\citet{king2013censorship} demonstrate that Chinese censors permit criticism of individual officials but systematically suppress content with collective action potential, leading activists to develop linguistic workarounds—homophonic substitutions, metaphors, and coded language that evade keyword filtering. "Crab culture" on Chinese social media denotes how activists use animal metaphors and visual symbols to discuss taboo topics~\cite{ng2021crab}. These studies establish that censorship pressure drives linguistic innovation, with activists exploiting language's flexibility to communicate despite restrictions.

Mandarin presents unique affordances for censorship evasion due to its morpho-syllabic writing system. A single sound can be represented by multiple characters with different meanings, and characters can be visually altered while remaining phonetically recognizable. This linguistic feature enables activists to communicate "in plain sight" while evading keyword filters that platforms deploy.

However, this literature focuses almost exclusively on state censorship in authoritarian regimes. Protestors increasingly face what~\citet{roberts2018censored} terms "corporate censorship"—content moderation by private platforms pursuing commercial or reputational objectives, deployed at a transnational level rather than being country-specific. As a democracy, this corporate censorship is precisely what Bluebird activists confronted.

\subsection{Threads and Taiwan}

Threads, launched by Meta (the parent company of Facebook and Instagram) on July 5, 2023, is a microblogging platform designed for sharing short-form text, multimedia content, and facilitating public conversation, tightly integrated with Instagram accounts and social graphs. Initially conceived under the internal codename ``Project~92,'' Threads rapidly positioned itself as a competitor to Twitter—now X—leveraging Meta’s existing infrastructure to gain 10 million users within its first seven hours and over 100 million sign-ups in its first five days, becoming the fastest-growing consumer app in history~\cite{meta2023threads,perelli2025meta,zahn2023threads}. Meta explicitly designed Threads as a platform for more engaging public discourse and creator–fan interactions, leveraging parasocial relationships—one-sided social relationships in which individuals feel a sense of intimacy with a public figure, even though the interaction is not reciprocal~\cite{giles2002parasocial,horton1956mass,tukachinsky2018theorizing}. As of August 2025, Threads reported 400 million users across more than 100 countries~\cite{brandonMetaSaysThreads2025}.

Taiwan accounts for an astounding 23.75\% of Threads’ global web traffic, the highest among all countries~\cite{asiapac2025threads}. Prior to Threads, Taiwan was dominated by a few social platforms. LINE, a chatroom app, is Taiwan’s most dominant social platform used by 94\% of Taiwan’s population. Facebook is used by 17.1 million users as of January 2025, representing 93.6\% of the population. Lastly, Instagram serves 11.3 million users in Taiwan, which corresponds to 49.1\%. These platforms previously formed the core of Taiwanese politics, with significant levels of discourse in the 2020 and 2024 Taiwanese elections~\cite{chang2024taiwanese,chang2021digital}.

In Taiwan’s lexicon, Threads quickly earned the humorous nickname ``cuì'' (meaning ``crispy'' in Mandarin). Notable political figures in Taiwan, including former President Chen Shui-bian, have used Threads in populist, civics-focused ways—like answering students’ legal questions. Due to its public-facing nature, Threads became the Twitter Taiwan never had. Threads became a new gathering space, especially for young users in Taiwan, to discuss political content, and eventually, political mobilization~\cite{yang2024threads}.

The general election of Taiwan, held in January 2024, featured a three-way horse race between the Democratic Progressive Party (DPP), the Kuomintang (KMT), and Taiwan People’s Party (TPP). Taiwan's main political cleavages center on identity and territory, unlike traditional left-right divisions, and social movements have directly shaped party formation \cite{nachman2025}. In the past, presidential elections in Taiwan were a bipartisan race between the DPP and KMT, split between cross-strait attitudes. Growing concern with issues such as energy policy and the domestic economy allowed third parties like the TPP to emerge~\cite{weener2025beyond}. Although President Lai Ching-te initiated a record third presidency for the DPP, the KMT and TPP held a majority in the Legislative Yuan (LY), equivalent to Congress in the United States. Through a majority alliance, the KMT–TPP coalition could thus effectively block initiatives by the DPP. Many citizens voiced concerns that if legislative voting was done solely on the basis of partisanship, then this LY would diminish Taiwan’s democracy.

The Bluebird Movement was a youth-led civic protest movement in Taiwan that erupted in May 2024 in response to legislative maneuvers by the opposition KMT and TPP~\cite{wang2024protesters}. The protests were triggered by proposed parliamentary reforms, which expanded the legislature’s investigatory powers in ways that could weaken checks and balances and concentrate power in the KMT–TPP coalition. Beginning on May 17, 2024, thousands of students and young people gathered outside the Legislative Yuan in Taipei. By June 4, 2024, the movement raised 2.6 million dollars to display an ad in Times Square~\cite{everington2024bluebird}.

As the Bluebird protests grew, activists called not only for the revocation of the reforms, but the recall of specific legislators who had championed them. This was seen by some as a constitutional remedy: if lawmakers had acted in ways that undermined checks and balances, the people could exercise democratic oversight through recalls. Building on the movement’s energy, civic groups—and later the DPP—launched what became known as the ``Great Recall'' campaign. This unprecedented effort sought the removal of up to 31 legislators, mostly from the Kuomintang (KMT), whom activists accused of obstructing democratic governance. 

\subsection{Research Questions}

The Bluebird movement provides a \textbf{unique vantage point} for understanding contemporary protest. It captured a rare case where a significant portion of a country migrates to an entirely new platform. In one of the few Mandarin-speaking democracies, the main language provides unique insights as to how linguistic features circumvent not government, but platform-based, censorship. Lastly, it provides insight regarding how AI is visually deployed to mobilize citizens. 

Our study thus focuses on the following interrelated research questions:
\begin{enumerate}
    \item \textbf{RQ1: }What is the divergence, if any, between exposure and engagement, and across partisanship?
    \item  \textbf{RQ2:} What textual and visual communication strategies predict engagement? 
    \item \textbf{RQ3:} How is generative AI deployed in protest and counter protest?
\end{enumerate}

\section{Methods}

\subsection{Data Collection}

\begin{figure*}[h]
    \centering
    \includegraphics[width=0.85\linewidth]{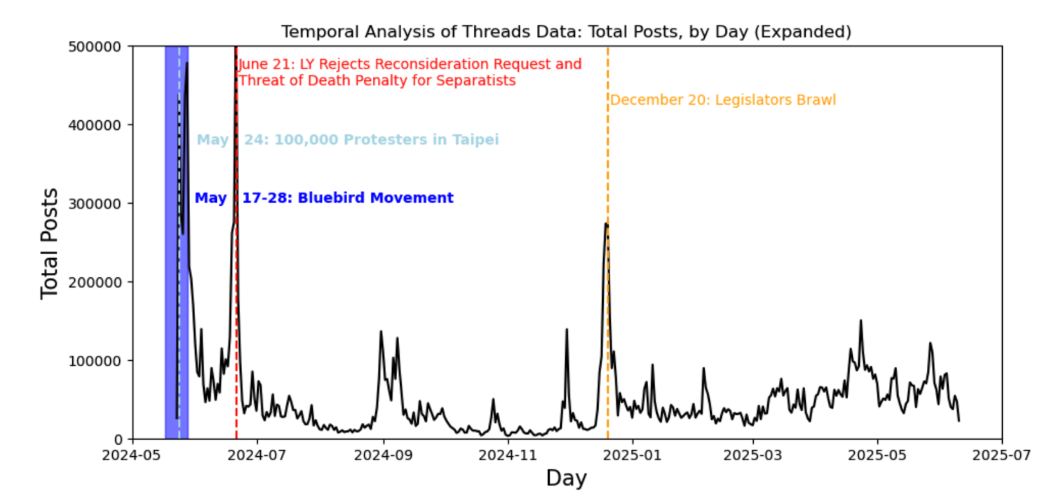}
    \caption{Annotated timeline of the Bluebird Movement between 05/01/2024 to 06/01/2025.}
    \label{fig:timeline}
\end{figure*}

We collected Threads posts related to the Bluebird Movement using the Meta Content Library (MCL) by saving the browsing sessions. We utilized only one keyword—Bluebird (qīngniǎo), and filtered for posts only containing traditional Chinese. This yielded a dataset of 62,321 unique posts from Threads between 05/01/2024 to 06/01/2025, capturing 13 months of the movement. Figure~\ref{fig:timeline} shows the time-series of the posts during this period. This data included the username, date, post text, and crucially, exposure and engagement metrics such as likes, comments, reposts, and views.

As a quick note on views, according to MCL’s documentation, views are the number of times the post appeared on screen, not counting the owner's own views. A post has no view count if there were fewer than 100 views. Additionally, research from Meta utilizing similar metrics define a view if a post is surfaced for greater than a certain visual proportion and number of seconds, which is meant to prevent overcounting from scrolling. As a test of quality, a random sample of 100 posts yielded 97\% directly related to Taiwan or social mobilization; the other 3\% were explicitly about mobilization outside of Taiwan, such as in Spain and France. Coupled with the fact that such a large amount of global traffic emerges from Taiwan and the unique geopolitical distribution of traditional Chinese (used only by Taiwan and Hong Kong, the latter for which no longer has Meta access), the quality of data is extremely high. However, it is unknown if there are foreign actors creating fake accounts; we discuss this as an avenue for additional investigation. Using the MCL urls, we then downloaded images associated with the posts, which yielded 21,572 unique photos. Not all posts have images, and some posts have multiple images.

\subsection{Data Augmentation with Language Models}

We rely on zero-shot learning to annotate posts without task-specific labeled data. In zero-shot setups, a model assigns labels it has never been explicitly trained on by leveraging shared semantic structure or broad pretraining. Early work framed this as mapping inputs and labels into a common semantic/attribute space to enable transfer~\cite{palatucci2009zero,socher2013zero,frome2013devise,xian2017zero}. 

In NLP, large pretrained language models exhibit strong zero-shot ability via instruction-following and prompt-based reformulations of classification (e.g., as cloze or NLI tasks), enabling accurate labeling from natural-language task descriptions alone~\cite{radford2019language,brown2020language,yin2019benchmarking,gao2021making,liu2023pretrain}. In our pipeline, we operationalized zero-shot labeling by prompting GPT-4o with a detailed schema and decision rules—without providing examples—and constraining outputs to a predefined JSON format.

\textbf{Text annotation.} For every post, the model extracted basic identifiers (languages, hashtags, locations, music references, dates/times, user tags, emojis). Building on existing typologies of protest communication, we also created a custom macro-set that included the following categories: Condemnation, Support, Personal Testimony, Humor/Satire, Call to Action, News Sharing, Cultural/Historical Context, Question/Information Seeking, Commemorative, Other. The model also flagged likely authorship (human vs.\ AI) and generated a $\leq$120-word summary. Tone and style were rated on 1–5 scales for sentiment toward the Bluebird Movement, formality, and aggressiveness. Content coding captured subjects; domestic vs.\ international politics; references to animals, students, politicians, influencers, and national-flag emojis. Domestic-issue indicators included death penalty, separatism, Legislative Yuan, occupation, recalls, misinformation, LGBTQ+, Indigenous rights, energy, budget, stock prices, electricity prices, overseas Taiwanese, healthcare, and democracy. Party variables recorded mentions of DPP/KMT/TPP, with 0–100 probabilities of supportive or oppositional stance toward each party. This prompt is provided in Appendix~A. 

\textbf{Image annotation.} For every image, the model (i) performed basic identification (typology, real vs.\ AI, artwork/cartoon status, short description, realism, meme likelihood, and a higher-level visual category); (ii) extracted any embedded text verbatim (OCR-style) and associated metadata (language(s), hashtags, locations, dates, links/QRs); (iii) coded visual content (objects and probabilities for blue birds, other animals and types, Taiwan/other maps, protests, references to Taiwanese/Hong Kong social movements, human presence, counts, signs, demographic descriptors, politicians, food, sexual content, merchandise, national flags by country, style cues such as kawaii/anime); (iv) coded political content (party references and rally likelihood), in-group/out-group partisan strategies (probabilities of support/attack by party); (v) cross-platform/media references (social platforms, news outlets, outlet language/type); (vi) alliances/international affairs (countries, Taiwan/ROC/China/PRC/CCP, Hong Kong, Milk Tea Alliance, Indo-Pacific/East/Southeast Asia, wars); and (vii) domestic issues (death penalty, Legislative Yuan, occupations, recalls, misinformation, LGBTQ+, Indigenous rights, energy). All probabilistic outputs were constrained to 0–100\% and all fields enforced via schema validation. We used deterministic decoding and strict JSON validation (see Appendix~A) and validations can be found in Appendix~B.

\subsection{Note on Data Ethics}

Access to Meta Content Library requires pre-approval and therefore has a data-sharing agreement. Additionally, while LLM labeling may generate bias, we validated the results manually to ensure suitable accuracy and no imbalance in false negatives and positives. Additionally, prior research has shown labeling elements directly, rather than vague themes, can improve reliability. 

Additionally, we were careful that our visualizations only include macro-aggregation of phenomenon, and the presented visuals are ones that appear multiple times. As a broader point, while our re-identification risk is low, replicability does suffer due to our inability to share the data directly. This is a trade-off commonly observed in modern data sharing agreements.

\subsection{Gradient Boosting Trees and SHAP Explainers}

Gradient boosting trees (GBTs) are an ensemble machine learning method that builds a predictive model in a stage-wise manner by sequentially fitting decision trees to the residuals of prior trees. At each iteration, a new tree is trained to approximate the negative gradient of the loss function with respect to the current model’s predictions, effectively performing functional gradient descent in function space~\cite{friedman2001greedy}. By iteratively adding these weak learners, GBTs combine the flexibility of decision trees with the power of boosting to capture complex, nonlinear relationships while maintaining strong generalization performance. Regularization techniques such as shrinkage (learning rate), subsampling, and tree depth constraints are commonly used to reduce overfitting and improve robustness~\cite{hastie2009elements}. 

Modern scalable implementations such as XGBoost~\cite{chen2016xgboost}, LightGBM~\cite{ke2017lightgbm}, and CatBoost~\cite{dorogush2018catboost} extend the original framework with efficient algorithms, parallelization, and categorical feature handling, making GBTs one of the most widely adopted methods for classification, regression, and ranking tasks across domains. This study uses CatBoost, which is well-suited for regressing with categorical variables through combinations of categorical features~\cite{prokhorenkova2018catboost}. We included a 75-25 train-test split, where the model was trained on 75\% of the data then validated on a 25\% random sample. The learning rate was set to 0.01 with a total of 1,500 epochs, with the tree depth set to 12. This was the result of a hyperparameters  grid search over tree depth (6, 9, 12), learning rate (0.03, 0.06), L2 regularization strength (3, 10), and Bayesian bootstrap settings with bagging temperatures of 0 and 2, to provide controlled stochasticity in sampling.

While machine learning techniques yield higher accuracies than canonical statistics, their “black box” nature has limited their interpretability for social science research. In recent years, SHAP explainers have become a useful tool for computing feature importance. Based on Shapley values in game theory, the idea is to calculate the utility contributions of each player to a coalition, based on different power sets of players~\cite{hart1987shapley}. Instead of players, SHAP evaluates power sets of features and their contribution to minimizing error in the model~\cite{lundberg2017unified}.

\section{Results}

We present results in three parts. First, we document the exposure-engagement gap across partisan frames, revealing how algorithmic visibility and user circulation operate according to distinct logics. Second, we identify textual strategies that predict virality, emphasizing commemorative and testimonial content. Third, we analyze visual communication patterns, with particular attention to AI-generated imagery as a tool for both mobilization and attack.

\subsection{Textual Analysis}

\begin{figure}[!htb]
    \centering
    \includegraphics[width=0.95\linewidth]{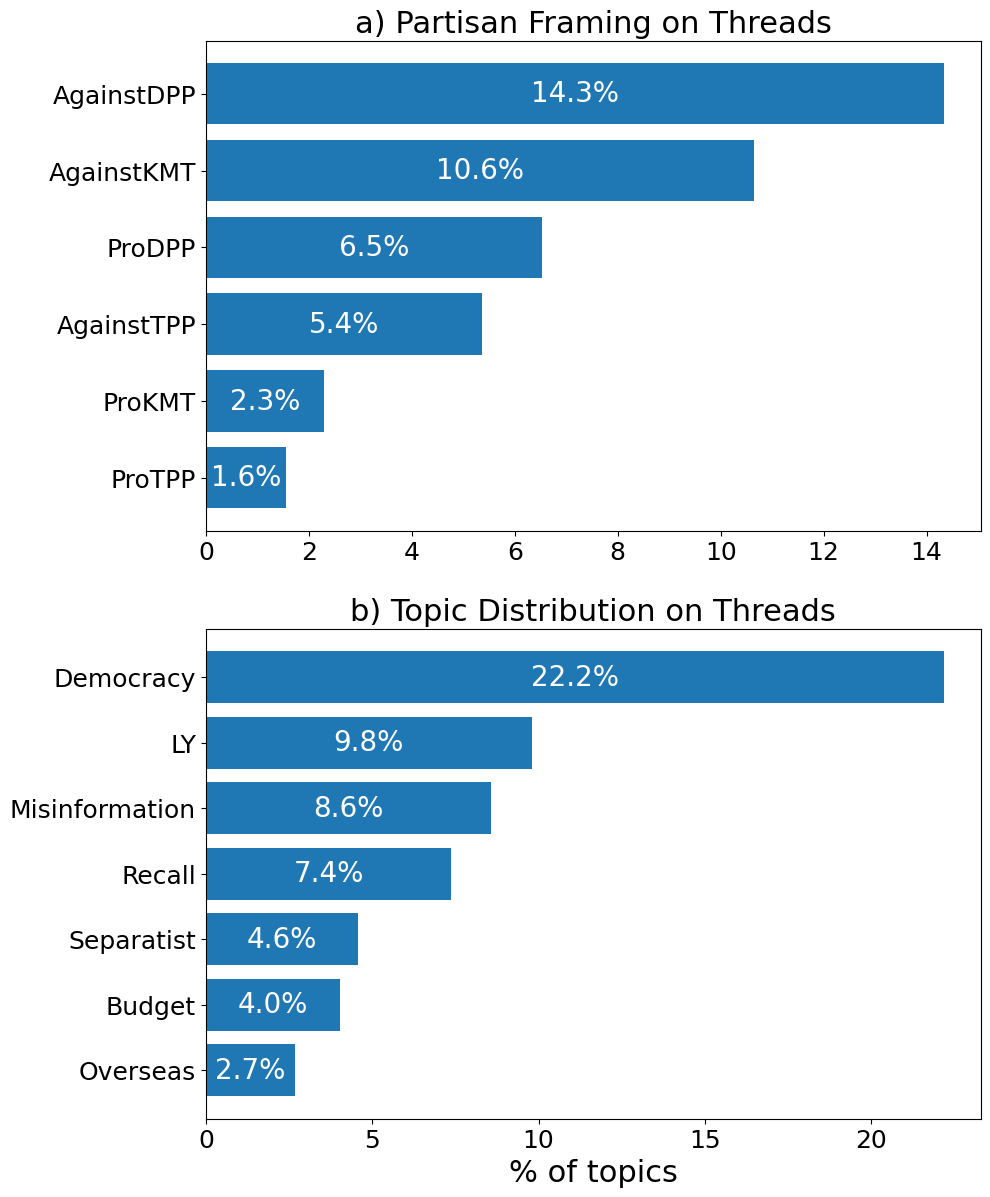}
    \caption{Textual proportion of Bluebird Threads content, with the (a) partisan framing  and (b) topical mentions within all posts.}
    \label{fig:textdist}
\end{figure}

Figure~\ref{fig:textdist}a shows the partisan framing of messages in the overall dataset. Analysis of 62,321 posts reveals partisan animosity as the dominant frame, with anti-DPP (14.3\%) and anti-KMT (10.6\%) content far exceeding supportive messaging. 
In contrast, only around 7\% of posts are pro-DPP and 6\% are anti-TPP. 

Turning to topics, Figure~\ref{fig:textdist}b shows the distribution of key topics in the overall dataset. By far, democracy is the most frequently evoked issue, covering more than one fifth of all content. The Legislative Yuan (LY, 9.8\%), references to misinformation (8.6\%), and the recall itself (7.4\%) take up the rest. These results have some immediate implications. While previous literature has shown that Threads is dominated by young people and DPP supporters, these results show that the supply of DPP-leaning content (i.e., anti-KMT and pro-DPP) is roughly in parity with KMT-leaning content.

\begin{figure}[!htb]
    \centering
    \includegraphics[width=1\linewidth]{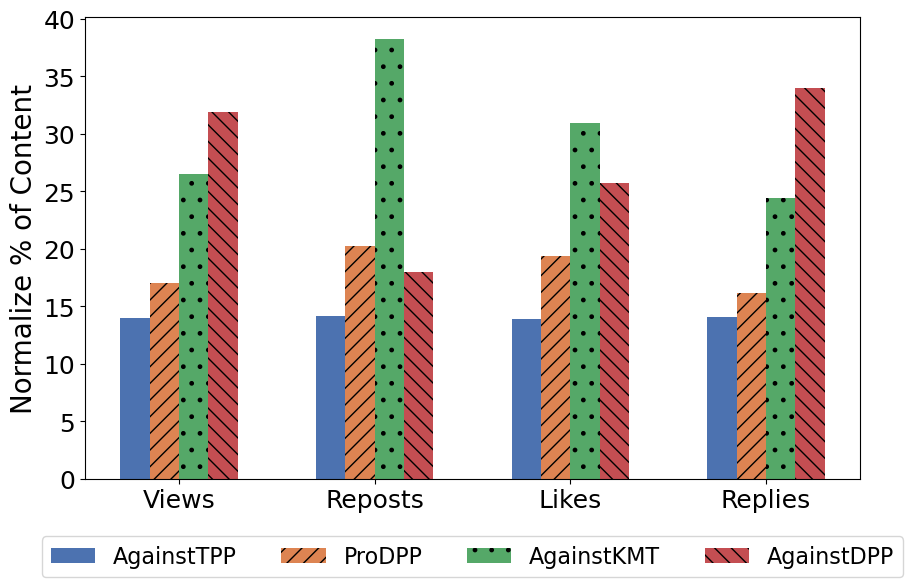}
    \caption{Top four partisan stances weighted by exposure (views) and engagement (reposts, likes, and replies).}
    \label{fig:partisanexposure}
\end{figure}

This partisan balance in content production makes the exposure-engagement divergence particularly striking, as differences in algorithmic visibility and user circulation cannot be attributed to underlying supply asymmetries---or content that is posted~\cite{zha2024genderinequalitiescontentcollaborations}. We further consider these stances based on key metrics in exposure and engagement. Figure~\ref{fig:partisanexposure} shows the top four partisan frames, weighted by their exposure and engagement then normalized against each other. For views, anti-DPP narratives are the most frequent, followed by anti-KMT narratives. However, for reposts, we observe a substantial decrease in anti-DPP reposts, relative to the number of anti-KMT reposts. This suggests that while the feed-ranking algorithm favors anti-DPP content, actual user engagement massively favors sharing anti-KMT content and does not favor anti-DPP content. This is corroborated by the increase in pro-DPP content, exceeding even anti-DPP content for reposts.

These patterns reveal a fundamental asymmetry. Anti-DPP content received 28\% more views than anti-KMT messaging despite comparable prevalence (14.3\% vs. 10.6\% of posts). However, anti-KMT content received 45\% more reposts. Pro-DPP frames similarly exceeded their view share in repost and like distributions. For replies, anti-DPP content attracted disproportionate activity, accounting for 34\% of all reply volume despite representing just 14\% of posts.

This divergence between exposure and engagement metrics reveals patterns in how content circulates. Views reflect what appears in users' feeds, while reposts and likes capture what users choose to amplify. Replies may signal contestation as much as endorsement. These metrics suggest the possibility of overlapping publics, driven by misalignment of the algorithm and users' intent. These answer \textbf{RQ1}.

\begin{figure*}[!htb]
    \centering
    \includegraphics[width=0.9\linewidth]{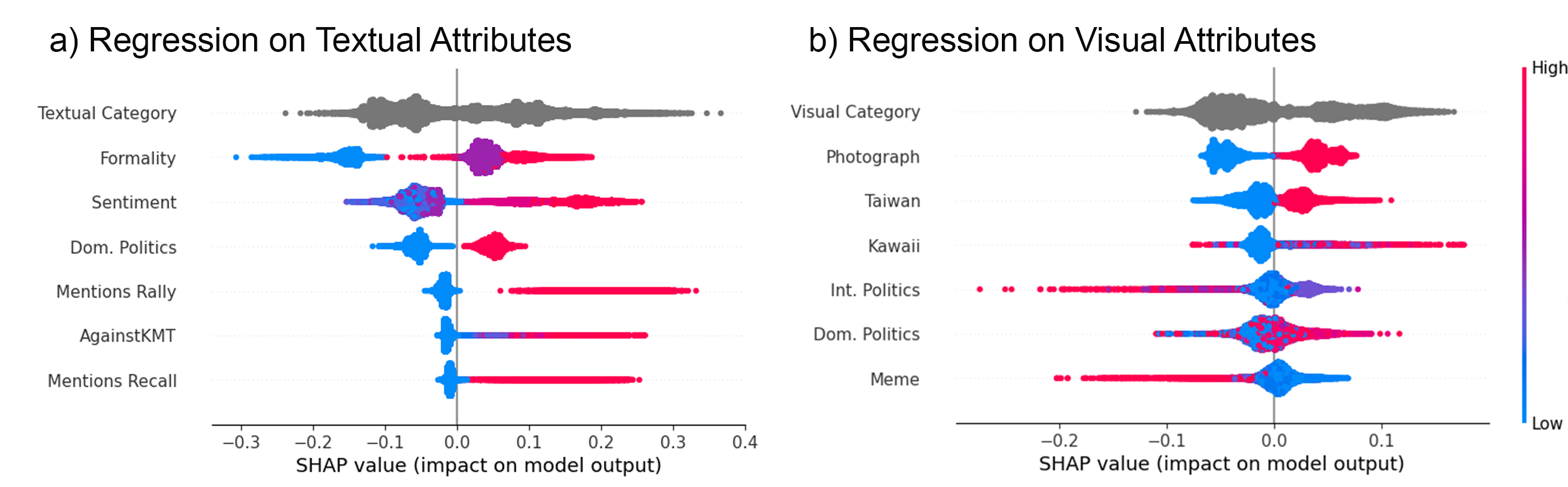}
    \caption{SHAP explainers of the CatBoost Regressor that predicts engagement (likes) based on a) textual and b) visual features.}
    \label{fig:shap}
\end{figure*}

Lastly, to identify the key variables that are driving virality, leveraging SHAP values. Figure~\ref{fig:shap}a) shows the SHAP values for the textual regressor. Being right (left) of the vertical line indicates a positive (negative) effect on likes; red (blue) dots correspond to high (low) levels for each specific variable. All variables along the y-axis are ranked according to their overall importance. The model yielded an $R^2$ of 0.42. Checks for colinearity (VIF) can be found in the Appendix Table~\ref{tab:vif_text}.

% \begin{figure}[!htb]
%     \centering
%     \includegraphics[width=1\linewidth]{AnonymousSubmission/LaTeX/figures/fig4.png}
%     \caption{SHAP explainers of the CatBoost Regressor that predicts likes based on 47 textual variables.}
%     \label{fig:shaptext}
% \end{figure}

The most important feature that predicts likes is the category of discussion. The top five category means include posts that are Commemorative (523.93), Personal Testimony (354.00), News Sharing (319.06), Calls to Action (311.26), and Cultural and Historical Context (305.38). Commemorative posts in particular mark protest milestones and reference historical parallels to the Sunflower Movement and invoked collective memory.
Other continuous variables show clear patterns: tonally, formal language and negative sentiment lead to more likes. For topic, posts related to political rallies, domestic politics, and the recall yield strong positive coefficients. Lastly, anti-partisan framing also yields the greatest virality.

\subsection{Visual Analysis}

Visual framing follows similar patterns. Figure~\ref{fig:shap} shows the SHAP values for a CatBoost model built on 93 visual elements. A few clear patterns emerge. Posts that contain actual photographs with textual or visual reference to Taiwan have a significant increase in virality. Similar to the textual case, images with reference to domestic politics yield more attention, whereas ones referencing international politics have generally less.  Checks for colinearity (VIF) can be found in the Appendix Table~\ref{tab:vif}. This answers \textbf{RQ2}.

\begin{figure}[!htb]
    \centering
    \includegraphics[width=1\linewidth]{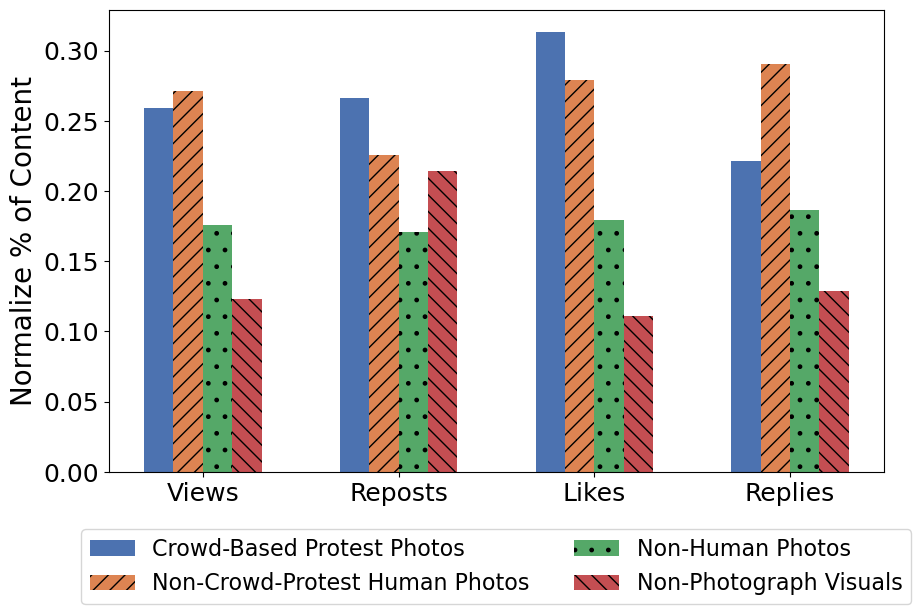}
    \caption{Attention funnel for visual framing categories across exposure and engagement metrics (views, reposts, likes, and replies).}
    \label{fig:attentionfunnel}
\end{figure}

Crucially, the visual category is the strongest predictor of likes. To examine this more closely, Figure~\ref{fig:attentionfunnel} shows the attention funnel of four visual categories. Overall, images with humans yield the greatest share of views, reposts, and likes, roughly evenly split between crowd-based and non-crowd human photos. This suggests direct references to humans yield greater exposure and engagement. 

However, a big shift emerges with replies, where non-crowd human photos yield a significant number of replies, whereas crowd-based protest photos yield significantly fewer replies. These human photos are likely pictures of legislators, which prompt fervent discussion, while protest crowd photos are viral but do not elicit discussion. Non-human photos have roughly the same share across the funnel. On the other hand, non-photograph visuals have a substantial boost in reposts, which suggests they are used as visuals for protest mobilization and information sharing.

\begin{figure}[!htb]
    \centering
    \includegraphics[width=1.1\linewidth]{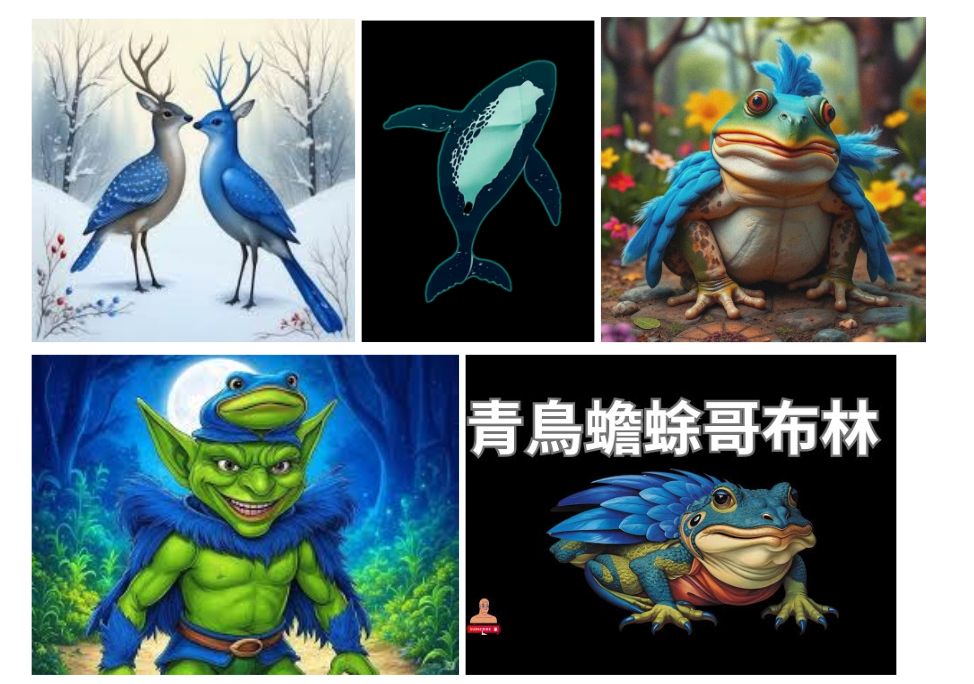}
    \caption{Samples of popular AI-generated images in the dataset, including a) blue bird and winter deer, b) whale island, c) a blue bird toad, and toad goblins for d) and e).}
    \label{fig:aiexamples}
\end{figure}

Diving deeper into the category of non-photograph visuals---content intended to mobilize---one interesting commonality is the large amount of animal and plant related imagery, created using generative AI. Table~\ref{tab:animalsymbols} summarizes the 10 different types of symbols utilized during protest discourse and Figure~\ref{fig:aiexamples} gives examples of AI-generated images. 

\begin{table*}[h]
    \centering
    \caption{Animal and plant imagery in the Bluebird Protests.}
    \label{tab:animalsymbols}
    \begin{tabular}{llll}
        \toprule
        Animal/plant & Translation & Origin & Political Orientation \\
        \midrule
        qīngniǎo & Blue bird & Homonym for Blue Island (qīngdǎo) street, to prevent suppression & Pro-DPP \\
        dōnglù & Winter deer & Homonym for East Road, part of protest location & Pro-DPP \\
        jīngdǎo & Whale & Homonym for Qing-Dao (qīngdǎo) street & Pro-DPP \\
        hēixióng & Black bear & Common symbol for Taiwanese nativism & Pro-DPP \\
        Lányīng & Blue eagle & Refers to KMT supporters (“eagles eating blue birds”) & Pro-KMT \\
        xiǎocǎo & Little grass & Refers to TPP supporters, from popular song & Pro-TPP \\
        Gēbùlín & Goblin & Used in harassment scandals; shifted to anti-DPP & Anti-DPP \\
        chánchú & Toad & Refers to DPP legislators supporting martial law & Anti-DPP \\
        Lài háma & Toad & Derogatory nickname for Lai Ching-te (sound pun) & Anti-DPP \\
        kǎoyā & Roast duck & Linked to Chinese identity; pro-China supporters & Anti-KMT \\
        \bottomrule
    \end{tabular}
\end{table*}

Pro-DPP symbols are typically derived via \emph{Xieyin geng}, or puns based on homophones, and explicitly translated to some type of animal. In particular, Blue Bird Winter Deer (qīngniǎo dōnglù) is a homophone for Qing-Dao East Road, the street where the first demonstration of the movement took place. Whale Island (jīngdǎo) is another homophone that was used, but less popular. Figure~\ref{fig:aiexamples}a shows an AI-generated image of a bird-deer hybrid, while Figure~\ref{fig:aiexamples}b shows a whale encircling Taiwan. These symbols are animals endemic to the country, most exemplified by the Formosan Black Bear. Although the black bear does not have a direct pun, it is often regarded as a symbol of Taiwanese nativism and identity.

However, AI-generated images can also be used to attack. Anti-DPP symbols include goblins and toads, shown in Figures~\ref{fig:aiexamples}c--e. Goblins were initially an anti-TPP term after alleged sexual harassment in the party. However, after DPP whistleblowers accused high-ranking members of the party of sexual harassment, the term evolved to become more associated with anti-DPP sentiment. This association was exacerbated by a television series inspired by the scandal.

The nickname ``Pro–Martial Law Toad'' originated from a post by the Democratic Progressive Party (DPP) caucus in Taiwan’s Legislative Yuan on social media. In that post, the DPP expressed support for South Korean President Yoon Suk-yeol’s declaration of martial law, describing it as a move to ``safeguard the system of freedom and constitutional governance.'' This statement sparked widespread controversy both domestically and internationally. In particular, users on South Korean online forums voiced strong dissatisfaction with the DPP’s stance, even going so far as to call the party a ``disgusting toad.'' This symbol had additional synergy as President Lai was also frequently called a toad by his political opponents, due to another name for toad sharing a word with his surname (i.e., Lai Hama).

In sum, the case of Taiwan demonstrates how generative AI can be used in modern social movements. Reposted more than it is liked or viewed, it serves as a way to mobilize. More crucially, it exemplifies a new paradigm: hiding toxic content using cute images, as another nefarious consequence of the aesthetics of cuteness~\cite{ngai2012our}. Moreover, textual and visual strategies converged on similar principles. High-virality content tends to reinforce collective identity through commemorative narratives and protest photos, or circulated as low-friction mobilization tools through personal testimony and AI-generated symbols. This answers \textbf{RQ3}.

\section{Discussion}

The Bluebird Movement illustrates how a newly emergent platform can become central to democratic contention. Our analysis of over 62,000 Threads posts reveals that the dynamics of protest communication on this platform cannot be reduced to the volume of content produced. Instead, the results show an asymmetry between what the algorithm amplifies and what citizens choose to endorse. Anti-DPP posts received greater exposure through the feed-ranking system, yet user engagement disproportionately elevated anti-KMT and pro-DPP content~\cite{chang2024taiwanese,bennett2012logic}. This suggests that platform logics and social networks may work at cross-purposes: while algorithms surface content that aligns with particular partisan attacks, publics recirculate frames that defend their own side or mobilize against opponents. Such a pattern resonates with recent work emphasizing the dual logics of algorithmic curation and connective action in networked protest~\cite{bennett2012logic,freelon2020false,tufekci2017twitter}. In this sense, Threads did not merely mirror offline cleavages but actively structured which political identities were rendered visible and which gained traction.

The exposure-engagement gap we document—anti-DPP content receiving 28\% more views but 45\% fewer reposts than anti-KMT content—points to the possibility of overlapping publics operating according to distinct logics~\cite{bandy2021more}. Views reflect what appears in users' feeds; reposts and likes capture what users choose to amplify. Replies may signal contestation as much as endorsement. Rather than a unified public sphere, these divergent metrics suggest parallel information environments where algorithmic curation and user intent do not align uniformly.

Equally important is the role of textual repertoires. Posts that commemorated events, shared personal testimony, or called others to action were the strongest predictors of virality. Rather than short bursts of humor or casual commentary, it was messages embedded in collective memory, affect, and civic duty that traveled farthest. This aligns with longstanding theories of moral shock and collective identity in social movements~\cite{mcadam2001dynamics,jasper1998emotions,gerbaudo2012tweets} and with empirical findings that moral-emotional language fuels diffusion on social media~\cite{brady2017emotion,chang2022justiceforgeorgefloyd}. Even within a platform optimized for brevity and casual conversation, protest actors crafted texts that linked personal voice to collective stakes. Formal tone and negative sentiment further amplified engagement, suggesting that moments of political threat are framed most effectively through seriousness rather than irony—a contrast to prior accounts that emphasize humor or satire in online activism~\cite{penney2017social}.

Visual communication deepened these dynamics in distinctive ways. Human photographs commanded the largest reach and sparked conversation, particularly when they depicted legislators. These images personalized accountability, directing attention toward elites and inviting direct confrontation in replies. By contrast, AI-generated imagery functioned less as a mode of deliberation and more as a vehicle for mobilization. Symbols such as blue birds, winter deer, and whale island—often punning on protest locations—circulated widely through reposts, providing a portable shorthand that reduced the costs of participation and strengthened collective identity~\cite{shifman2014memes}. Yet these same techniques were redeployed for attack: goblins, toads, and other AI-generated figures became tools of ridicule aimed at rival parties.

The dual use of symbolic AI imagery—mobilizing solidarity on the one hand and cloaking toxicity in cuteness on the other—demonstrates both the promise and the peril of generative aesthetics in contentious politics. This echoes scholarship on kawaii as a political aesthetic~\cite{allison2006millennial,ngai2012our} and extends recent work on meme culture and digital symbolism in protest movements~\cite{shifman2014memes,milner2016world}. In other words, we identify a new paradigm for out-group attacks—by packaging extreme toxicity in cute animals.

These findings carry broader implications for how digital protest repertoires evolve. Rather than displacing institutional channels, Threads communication was tightly coupled to them. The textual and visual strategies that spread most widely were those that connected online affect to offline mechanisms of accountability, particularly the unprecedented wave of recall elections. The Bluebird Movement thus shows how a new platform can bootstrap existing constitutional remedies, transforming symbolic protest into formal oversight. This dynamic illustrates the continued importance of institutional context in shaping the trajectories of digital mobilization~\cite{earl2011digitally,theocharis2015using}.

\section{Acknowledgments}
This research was supported by the Hanlon Scholars Program and Burke Research Initiative. We thank Hsuan Lo and Austin Wang for feedback.

\bibliography{aaai2026}

\clearpage

\newpage

\appendix

\section{Prompt for visual annotation}

The full prompt can be found in the GitHub.

\lstset{mathescape=true}
\begin{lstlisting}[breaklines=true, basicstyle=\small\ttfamily]
As a political respondent analyzing the domestic politics of Taiwan and international politics, your goal is to assess and provide accurate descriptions of images.

### Instructions

The following formatting notes are extremely important to follow exactly correctly:
- If an image includes any text, please extract it all into a string.
- Please give probabilities as percentage likelihood (i.e. 1% if very unlikely and 99% if extremely likely).
- Corresponding to the questions below, you will need to output a JSON object.
- Return the structured JSON only, with no additional text, descriptions, or explanations.

### Questions

This is an image that I want to upload. Describe what is in it.

=== Basic Identification ===s
- Classify the image into one of the following typologies: 
  (Map, Logo, Infographic, Weather Map, News Footage, Legislative Footage, Merchandise, Art, Cartoon, Personal, Nature). If none fit, use "other". (label as Category:)
- Create a description of the image, limited to 60 words. (label as Description:)
- Classify the image into one of the following visual content categories:
  (Crowd-Based Protest Photos, Non-Crowd-Protest Human Photos, Non-Human Photos, Non-Photograph Visuals). 

=== Textual Content ===
- If the image includes any text, extract it into a string. (label as Text:)
- If it contains text, what language is it? If multiple languages, list all. (label as Language:)
- If the image includes any hashtags, extract them. (label as Hashtags:)
- If the image includes any locations, extract them. (label as Location:)

=== Visual Content ===
- What is the probability the image contains a blue bird? (label as BlueBird:)
- What is the probability the image contains animals other than blue birds? (label as Animal:)
- If animals are present, what kind? (label as AnimalType:)

=== In-Group/Out-Group Partisan Strategies ===
- What is the probability the image supports the DPP? (label as ProDPP:)
- Attacks the DPP? (label as AgainstDPP:)
- Supports the KMT? (label as ProKMT:)
- Attacks the KMT? (label as AgainstKMT:)

{
  "Category": string, // One of: Map, Logo, Infographic, Weather Map, News Footage, Legislative Footage, Merchandise, Art, Cartoon, Personal, Nature, other
  "Description": string, // $\leq$ 20-word description
  "VisualCategory": string, // One of: Crowd-Based Protest Photos, Non-Crowd-Protest Human Photos, Non-Human Photos, Non-Photograph Visuals, other

  "Text": string, // Full extracted text from image, if any
  "Language": [string], // List of languages (e.g., ["zh", "en"])
  "Hashtags": string,
  "Location": string,

  "BlueBird": int, // 0-100 (%)
  "BlueBirdFly": int,
  "Animal": int,
  "AnimalType": string,

  "ProDPP": int,
  "AgainstDPP": int,
  "ProKMT": int,
  "AgainstKMT": int,

}
\end{lstlisting}

\newpage

\section{Accuracy for stratified sampled labeling}

% Accuracy checks can be found at the
% \href{https://osf.io/j5e3q/overview?view_only=11ee872ae1ff4b90801ee38809dcc356}%
% {\textcolor{blue}{Open Science Framework files for this project}}.

% Accuracy checks can be found in the Open Science Framework files for this project:
% \url{https://osf.io/j5e3q/overview?view\_only=11ee872ae1ff4b90801ee38809dcc356}

Full accuracy scores can be found in the Open Science Framework files for this project:
\path{https://osf.io/j5e3q/overview?view_only=11ee872ae1ff4b90801ee38809dcc356}.

\begin{table}[!htb]
\centering
\caption{Model feature performance summary.}
\label{tab:feature_perf}
\begin{tabular}{@{}ll@{}}
\toprule
\textbf{Feature} & \textbf{Value} \\
\midrule
VisualCategory        & 91.9\% \\
Photograph            & 95.2\% \\
Taiwan                & 78.6\% \\
Cute                  & 100\% \\
InternationalPolitics & 92.1\% \\
DomesticPolitics      & 90.5\% \\
Meme                  & 100\% \\
\bottomrule
\end{tabular}
\end{table}

%\clearpage
%\newpage
\section{Checks for Colinearity}

\begin{table}[!h]
\centering
\caption{Variance Inflation Factors (VIF) for Visual-Based Predictors.}
\label{tab:vif}
\begin{tabular}{lc}
\hline
\textbf{Variable} & \textbf{VIF} \\
\hline
Intercept                & 5.18 \\
Photograph               & 1.06 \\
Taiwan                   & 1.88 \\
Cute                     & 1.22 \\
International Politics   & 1.33 \\
Domestic Politics        & 2.23 \\
Meme                     & 1.18 \\
Human Numbers            & 1.00 \\
DPP                      & 1.23 \\
KMT                      & 1.04 \\
TPP                      & 1.02 \\
\hline
\end{tabular}
\end{table}

\begin{table}[!h]
\centering
\caption{Variance Inflation Factors (VIF) for Text-Based Predictors.}
\label{tab:vif_text}
\begin{tabular}{lc}
\hline
\textbf{Variable} & \textbf{VIF} \\
\hline
Intercept                 & 12.00 \\
Formality                 & 1.23 \\
Sentiment                 & 1.30 \\
Domestic Politics (Text)  & 1.51 \\
Political Rally (Text)    & 1.14 \\
Against KMT               & 1.33 \\
Recall (Text)             & 1.11 \\
Pro-DPP                   & 1.16 \\
Against DPP               & 1.30 \\
Pro-KMT                   & 1.07 \\
Pro-TPP                   & 1.05 \\
Against TPP               & 1.15 \\
\hline
\end{tabular}
\end{table}

\end{document}